\documentclass{jfm}
\usepackage{graphicx}
\usepackage{epstopdf, epsfig}
\usepackage{amsmath}
\usepackage{tikz}
\usepackage{physics}
\usepackage{graphicx}
\usepackage{caption}
\usepackage{subcaption}

\def\BV{Brunt-V{\"a}is{\"a}l{\"a}\ }

\shorttitle{Damping of Quasi-2D internal wave attractors by rigid-wall friction}
\shortauthor{F. Beckebanze,
  C. Brouzet,  
  I. N. Sibgatullin and L. R. M. Maas}

\title{Damping of quasi-2D internal wave attractors by rigid-wall friction}

\author{F. Beckebanze\aff{1}
  \corresp{\email{f.beckebanze@uu.nl}},
  C. Brouzet\aff{2},  
  I. N. Sibgatullin\aff{3,4,5} \\
  \and L. R. M. Maas\aff{6}}

\affiliation{\aff{1} Mathematical Institute, Utrecht University, P.O. Box 80010, 3508 TA Utrecht, The Netherlands
\aff{2} Univ Lyon, ENS de Lyon, Univ Claude Bernard, CNRS, Laboratoire de Physique, F-69342 Lyon, France
\aff{3} Department of Mechanics and Mathematics, Moscow State University, Moscow 119191, Russia
\aff{4} Institute for System Prorgamming, Moscow, 109004, Russia
\aff{5} Shirshov Institute of Oceanology, Moscow, 117997, Russia
\aff{6}Institute for Marine and Atmospheric research Utrecht (IMAU), Utrecht University, Princetonplein 5, 3584 CC Utrecht, The Netherlands}

\begin{document}

\maketitle
\begin{abstract}
The reflection of internal gravity waves at sloping boundaries leads to focusing or defocusing. In closed domains, focusing typically dominates and projects the wave energy onto 'wave attractors'. For small-amplitude internal waves, the projection of energy onto higher wave numbers by geometric focusing can be balanced by viscous dissipation at high wave numbers. Contrary to what was previously suggested, viscous dissipation in interior shear layers may not be sufficient to explain the experiments on wave attractors in the classical quasi-2D trapezoidal laboratory set-ups. Applying standard boundary layer theory, we provide an elaborate description of the viscous dissipation in the interior shear layer, as well as at the rigid boundaries. Our analysis shows that even if the thin lateral Stokes boundary layers consist of no more than 1\% of the wall-to-wall distance, dissipation by lateral walls dominates at intermediate wave numbers.  Our extended model for the spectrum of 3D wave attractors in equilibrium closes the gap between observations and theory by \cite{Ha08}. 
\end{abstract}


\section{Introduction}
The dispersion relation of internal waves is given by
$\omega_0^2=N_0^2 \sin^2 \theta,$
with $\omega_0$ the wave frequency, $\theta$ the angle of phase propagation with respect to the vertical, $z$, antiparallel to gravity, and $N_0$ the \BV frequency, here assumed constant. 
The group propagation is always orthogonal to the phase propagation \citep{Su10}, thus $\theta$ also represents the angle of energy propagation with respect to the horizontal plane, and is fixed for monochromatic waves. This property results in geometric focusing or defocusing upon reflection at sloping topography. Repeated geometric focusing in closed domains can project the wave energy onto closed orbits, known as wave attractors \citep{ML95,Ma97}. In the vicinity of internal wave attractors, energy is dissipated by viscous dissipation \citep{Ha08}, or lost to nonlinear wave-wave interactions \citep{Sc13, Br16a, Br17, Da17}. Internal wave attractors are studied most thoroughly in the classical quasi-2D trapezoidal set-ups \citep{ML95,Ma97, Ma05, Ma09, SSM07, Harlander2008, Ha08, Ha10a, Gr08, Sc13, Br16a,Br16,Br17}, geometries which are also popular in studies on closely related inertia wave attractors \citep{Manders2003, Kl14, Tr17}. Recent studies also examine internal wave attractors confined to more sophisticated domains, resembling simplified ocean topography \citep{Ta10, Ec11, Ha11, Wa15, Gu16}. Applying standard boundary layer theory, \cite{Kl14} establish the importance of the Ekman boundary layers for inertial wave attractors. Surprisingly, the role of energy dissipation at rigid boundaries  for internal wave attractors still remains an open question, even for the simplest domain, the classical quasi-2D trapezoid. The energy loss at the wave attractor - and in the broader sense internal wave beams - can have far-reaching consequences for the mixing budget of stratified fluids, such as the deep oceans \citep{WF04} and marginal seas \citep{La14}. \\
In this paper, we apply standard boundary layer theory to quantify the frictional damping mechanisms of internal wave attractors in the classical quasi-2D laboratory set-up. Frictional dissipation takes place in two types of viscous layers: shear layers in the interior along the attractor and boundary layers at the rigid boundaries.  
\\
Internal wave damping through interior shear layers, first described by \cite{TS73}, has been studied extensively over the past decades, and in particular in the context of internal wave attractors by \cite{Di99, Sw07,Ha08, Br16a} and inertial wave attractors by \cite{Di99, R01,R02, Og05, Jo14}. A simple model for an equilibrium wave attractor spectrum, with the energy input at the basin scale (= low wave numbers) and dissipation only through internal shear at high wave numbers, has been derived by \cite{Ha08}. Although the structure of their theoretical spectrum resembles their experimentally observed spectrum of an internal wave attractor in the classical quasi-2D trapezoidal set-up, the discrepancy hints at significant dissipation at the rigid boundaries. \cite{Gr08} performed 2D numerical simulations, designed to replicate the laboratory experiment by \cite{Ha08} with free-slip boundaries. Their simulations underestimates the energy dissipation at high wave numbers, also indicating an additional energy sink at the walls in the laboratory. The fully 3D simulations by \cite{Br16} signify significantly increased dissipation rates in the lateral boundary layers. Our theoretical analysis shows that adding dissipation at the rigid boundaries closes the gap between the model and observations in \cite{Ha08}. 
\\
Stokes boundary layers in homogenous fluids are well-understood and are described in many text books on fluid mechanics, e.g. \cite{SG00}.
The stratified boundary layers for monochromatic internal waves are to some extend analogous to homogenous Stokes boundary layers, but differ on a number of fundamental aspects, such as the characteristic thickness of the boundary layer. The thickness of the stratified boundary layer is given by
$$d_0=\mu^{-1} \left(  \frac{\nu }{ \omega}\right)^\frac{1}{2}, \qquad \mbox{ with }\quad \mu =\sqrt{ \abs{ \frac{\sin^2 \varphi}{\sin^2 \theta}-1  }}$$
dependent on the angle $\varphi$ of the boundary (with respect to the horizontal) and the internal wave inclination, $\theta$. Note that horizontal boundaries ($\varphi=0$) coincide with the homogeneous case, $\mu=1$. For near-critical reflections ($\varphi \sim \theta$) the boundary layer thickness $d_0$ tends to infinity, making different approaches, such as in \cite{DY99}, necessary. The theoretical investigation on stratified rotating boundary layers by \cite{Sw10} stresses the importance of these critical cases ($\varphi \sim \theta$) for the generation of internal inertia waves by oscillating boundaries. \cite{KC95b,KC95a} computed the boundary layer of a reflecting internal wave beam, but did not account for the dissipative energy loss inside the boundary layer. \cite{VC03} constructed asymptotic solutions for internal wave fields generated by a rigid plane vibrating along its surface. We now investigate a situation in which the energy flux is in opposite direction, i.e. the wave attractor looses energy to the rigid walls. The objective is to understand and quantify the damping induced by stratified boundary layers on wave attractors. Partial results are also reported in \cite{Be16a}. 
 \\ 
The structure of this paper is as follows. The formulation of the problem is described in \S \ref{preliminaries}. In \S \ref{interior}, we construct inviscid wave attractor solutions. Internal shear, lateral wall boundary layers and boundary layers at the reflecting walls are subsequently added in respectively \S \ref{ivd}, \S \ref{lateral} and \S \ref{refl}. In \S \ref{comparison}, we compare our extended model for the equilibrium wave attractor spectrum with the laboratory experiment and 3D simulations.  Conclusions are drawn in \S \ref{conclusions}.

\section{Preliminaries}
\label{preliminaries}
\begin{figure}
\centering
\begin{minipage}{.5\textwidth}
 \hspace*{0.8cm}
  \includegraphics[type=pdf,ext=.pdf,read=.pdf, width=0.8\textwidth]{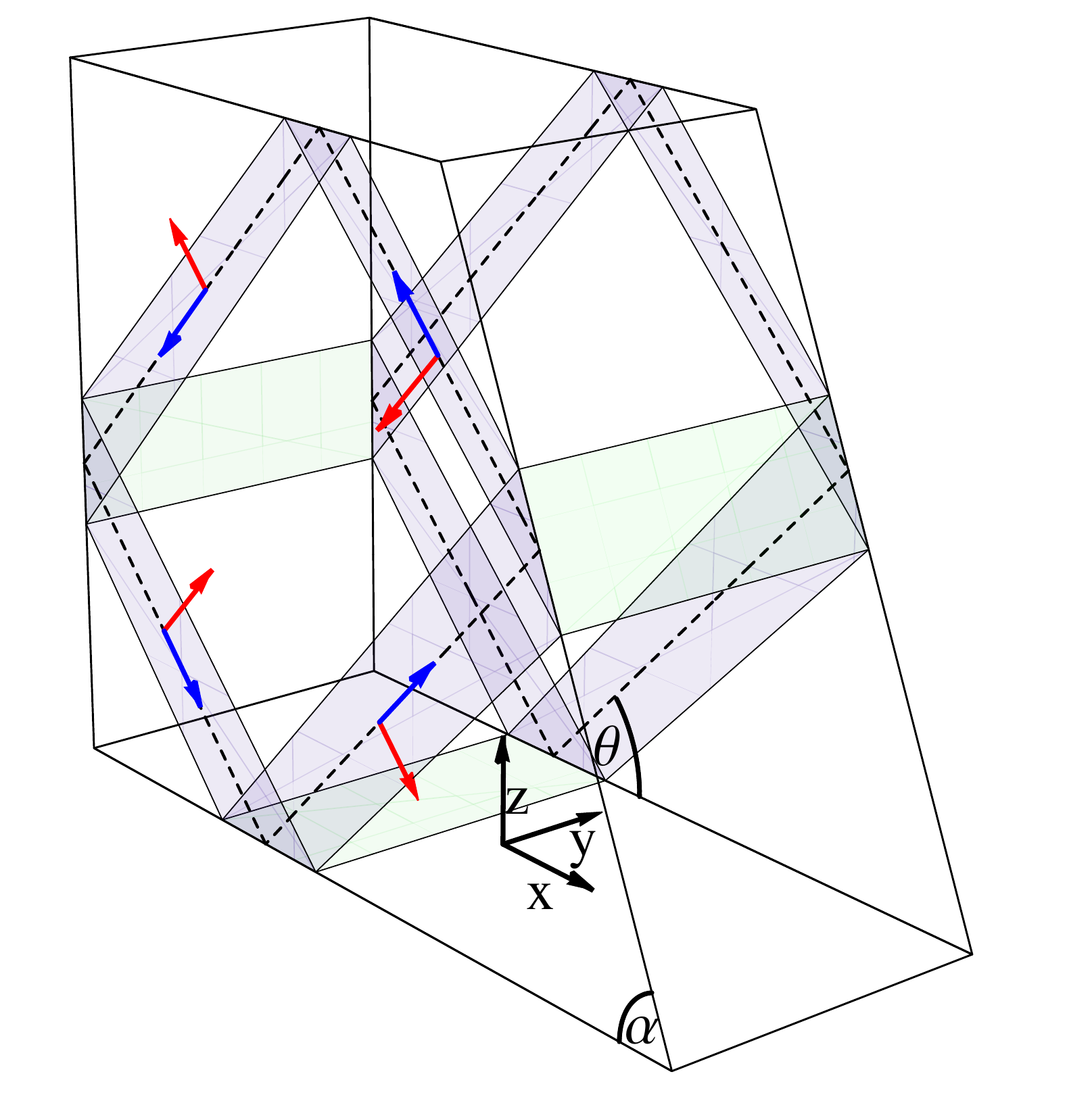}  
  \put(-165,140){(a)}
\end{minipage}%
\begin{minipage}{.5\textwidth}
  \hspace*{0cm}
  \includegraphics[type=pdf,ext=.pdf,read=.pdf, width=0.9\textwidth]{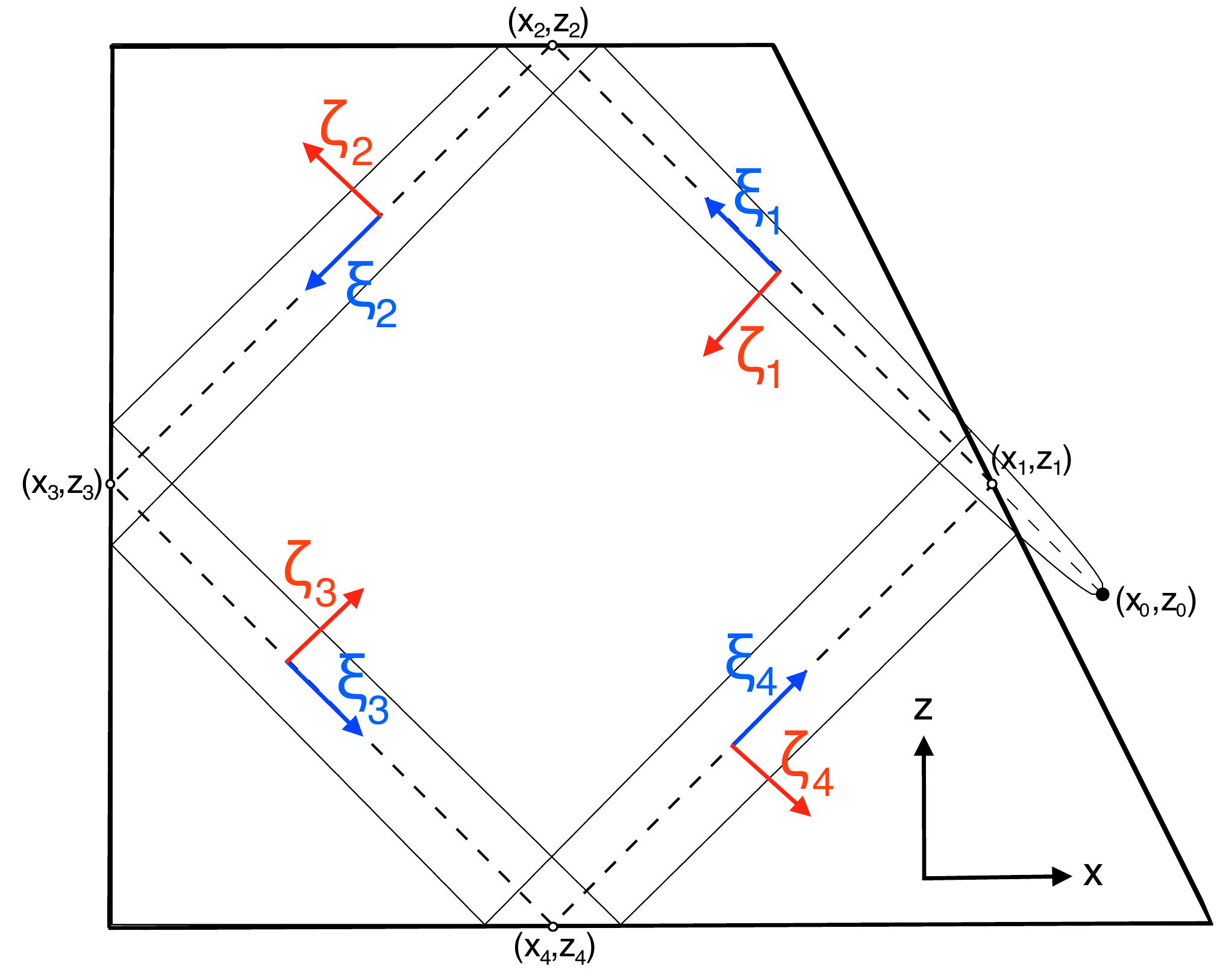}
  \put(-175,125){(b)}
\end{minipage}
\caption{Schematic view of the 3D trapezoidal domain $D$ (a), and its side view (b).  The viscous wave attractor interacts with the rigid boundaries of $D$ (free-slip at surface, $z=h $) primarily in the shaded areas (a), as it is confined to a region around the theoretical inviscid wave attractor orbit (dashed lines in a,b). The black dot in (b) to the right of the inclined wall indicates the virtual source. The phase propagation is along coordinates $\zeta_n$, $n=1,2,3,4$ (red arrows), whereas the energy propagation is along $\xi_n$ (blue arrows) at an angle $\theta$ with respect to the horizontal $x$. The thickening of the wave attractor in the energy propagation direction is due to viscous damping, as discussed in \S \ref{lateral}, and it is balanced by the focusing reflection at inclined wall with angle $\alpha$.   }
\label{f1}
\end{figure}

In this paper we consider monochromatic internal waves in a linearly stratified Boussinesq fluid inside a trapezoidal tank 
$$D=\{(x,y,z)\in \mathbb{R} \ \ | -l_y \leq y \leq l_y, \ \ - l_x \leq x\leq l_x, \ \ 0 \leq z \leq \min[ h,  (l_x-x) \tan \alpha]  \},$$
with $z$ antiparallel to gravity. We anticipate ratios $\sin \theta = \omega_0/N_0$ of wave frequency $\omega_0$ over \BV frequency $N_0$, such that the internal wave motion is predominantly confined to a neighborhood around the theoretical inviscid wave attractor, as illustrated in Fig. \ref{f1}a,b. The Cartesian coordinates, $(x,y,z)$, are dimensionalized with the length scale $L_0$, which we assume to be the characteristic wave length of the predominant wave motion - the viscous wave attractor - measured in the cross-beam direction. 
Note that scaling the non-dimensional half-bottom-length, $l_x$, half-width, $l_y$, and height, $h$, with the same length scale, $L_0$, leaves the angle of the inclined wall, $\alpha$, and the energy propagation angle, $\theta$, both with respect to the horizontal, invariant. We require $l_y\gtrapprox  1$, i.e. the dimensional width, $W=2l_y L_0$, is at least of the same order of magnitude as the wave attractor cross-beam length scale, $L_0$.  \\ 
We consider sufficiently weak monochromatic forcing, generating only small-amplitude wave motion. This means that the Stokes number, $\frac{U_0}{\omega_0 L_0}$, with $U_0$ the dimensional scale of the internal wave velocity, is small such that all non-linear advection terms can be neglected. \\
Under these assumptions, the (linearized) equations governing the dimensionless velocity field ${\bf u}=(u,v,w)$, buoyancy $b$, and pressure $p$ of the Boussinesq fluid, with scaled \BV frequency $N=N_0/\omega_0=\pm 1/\sin \theta$, are given in subscript-derivative notation by 
\begin{equation}
\begin{split}
{\bf u}_t=\,& - \nabla p+b {\bf \hat{z}}+\delta^2 \Delta {\bf u}+{\bf f}e^{-i  t}, \qquad b_t=  -N^2 w, \qquad \nabla  \cdot {\bf u}= 0.
\label{ge1}
\end{split}
\end{equation}
Here, $\delta=\frac{d_0}{L_0}\ll 1$ is the non-dimensional Stokes boundary layer width, with $d_0=\sqrt{\nu / \omega_0}$, and $\nu$ the dynamical viscosity constant. The forcing ${\bf f}={\bf f}(x,z)$ is assumed to be uniform in the transversal $y$-direction. For mathematical convenience, we consider ${\bf f}$ to be a localized source, located outside the trapezoidal domain, $D$, as illustrated in Fig. \ref{f1}b. This enables us to describe the viscous wave attractor as four branches of a viscous internal wave beam \citep{Og05}. The downside of this approach is a slight violation of the impermeability boundary condition at the inclined wall near the wave attractor upon incorporating viscous attenuation. We accept this disadvantage, which also underlies the theoretical 2D spectra by \cite{Ha08}, because it is irrelevant for the energy loss through the boundary layers of a quasi-2D weakly viscous wave attractor - the main objective of the presented analysis. \\

We solve the governing equations (\ref{ge1}) asymptotically with no-slip boundary conditions, ${\bf u}={\bf 0}$, at the boundary of the trapezoidal domain $D$ (except at the free surface, $z=h $, where we impose free-slip), by expanding the velocity vector ${\bf u}$ in the small parameter $\delta$, 
$$ {\bf u}={\bf u}_0+\delta{\bf u}_{1}+\mathcal{O}(\delta^2) ,$$
 and similarly for buoyancy $b$ and pressure $p$.  We start in \S \ref{interior} by solving (\ref{ge1}) at $\mathcal{O}(\delta^0)$ with free-slip boundary conditions for ${\bf u_0}$. Free-slip means that we only require the impermeability boundary condition to hold. Viscous attenuation is added in \S \ref{ivd}, and in \S \ref{lateral}, we extend ${\bf u}_0$ such that it vanishes at the lateral walls (surfaces along dashed theoretical attractor in Fig. \ref{f1}a, blue online). In \S \ref{refl}, we add correction terms in order to satisfy the no-slip boundary condition also at the reflection sides (green surfaces in Fig. \ref{f1}a, green online). 
 
\section{Wave attractor branches in interior}
\label{interior} 

It is convenient to express the four wave attractor branches in the rotated and shifted coordinates $[\xi_n,\zeta_n]$, $n=1,2,3,4$, given by 
 $$
 \begin{bmatrix}
         \xi_{1,3}\\
       \zeta_{1,3}
     \end{bmatrix}=    \mp \begin{bmatrix}
       \cos \theta \  - \sin \theta   \\
      \sin \theta  \ \ \ \ \cos \theta
     \end{bmatrix}   \begin{bmatrix}
       x-x_{1,3} \\
       z-z_{1,3}
            \end{bmatrix},   \quad 
            \begin{bmatrix}
         \xi_{2,4}\\
       \zeta_{2,4}
     \end{bmatrix}=    \mp 
     \begin{bmatrix}
       \cos \theta \ \ \  \ \sin \theta   \\
      \sin \theta  \  - \cos \theta
     \end{bmatrix}   \begin{bmatrix}
       x-x_{2,4} \\
       z-z_{2,4}
            \end{bmatrix},$$
with $[x_n,z_n]$ the reflection points of the attractor, see Fig. \ref{f1}b. 
The theoretical inviscid wave attractor (dashed lines in Fig. \ref{f1}) corresponds to $\zeta_n=0$, for $n=1,2,3,4$. \\
The inviscid $\mathcal{O}(\delta^0)$-velocity field generated by a monochromatic, localized source, describing the first wave attractor branch (labeled with super-script $1$), can be written as
\begin{equation}
{\bf u}_0^{[1]}={\bf \hat{\xi}}_1 U(\zeta_1)=[-\cos \theta, 0, \sin \theta] U(\zeta_1), \qquad   U(\zeta)= \int\limits_0^{\infty} \hat{U}(k) \exp \left[ i(k \zeta-t) \right] dk, 
\label{U1}
\end{equation}
where $k$ is the non-dimensional wave number (scaled by $L_0^{-1}$). Physical quantities are always the real part of the presented expression, and the hat on a coordinate always denotes the unit vector pointing in the direction of this coordinate, i.e. ${\bf \hat{\xi}}_1$ is the unit vector along the first wave attractor branch. The Fourier spectrum 
$$\hat{U}(k)=\frac{1}{2\pi} \int_{-\infty}^{\infty} U(\zeta) \exp[-i \zeta k] d\zeta$$
 of the along-wave-beam velocity component $U$ depends on the unspecified localized source at $(x_0,z_0)$. The main objective of the presented analysis is to derive constraints for $\hat{U}(k)$, based on geometric wave focusing (this section), and viscous dissipation (\S \ref{ivd} - \S \ref{refl}). 
 Note that $\hat{U}(k)=0$ for $k\leq0$ because no energy can propagate towards the source (by assumption).\\ 
 Subsequent free-slip reflections  of the first wave attractor branch at the surface, $z=h $, at the vertical wall, $x=-l_x$, and at the bottom, $z=0$, lead to the following velocity fields for the second, third and fourth wave attractor branches:
  \begin{equation}
 \begin{split}
& 			 {\bf u}_0^{[n]}={\bf \hat{\xi}}_n U(\zeta_n)\quad \mbox{for } \ n=2, 3, 4. 
  \label{U234}
 \end{split}
 \end{equation}
 The fourth branch returns to the inclined wall, $z=(l_x-x) \tan \alpha$, where the free-slip boundary condition reads 
     \begin{equation}
 \begin{split}
 \Re \left[ \left({\bf u}_0^{[1]}+{\bf u}_0^{[4]} \right) \cdot {\bf \hat{ n}}_{\alpha} \right]=0,
 \label{impermeability}
       \end{split}
 \end{equation}
 with ${\bf \hat{ n}}_{\alpha}= [\sin \alpha, 0, \cos \alpha]$ a normal vector of the inclined wall. On the inclined wall, $z=(l_x-x) \tan \alpha$, we have
     \begin{equation}
 \begin{split}
   {\bf u}_0^{[1]}  \cdot {\bf \hat{ n}}_{\alpha}=  -\sin \left[ \alpha -\theta \right] \int\limits_0^{\infty} \hat{U}(k)      \exp \left[i k\frac{ \sin [\alpha-\theta]}{\cos \alpha} \left(x-x_1 \right)   -it \right] dk 
   \label{u01na}
      \end{split}
 \end{equation}
 and
     \begin{equation}
 \begin{split}
   {\bf u}_0^{[4]}  \cdot {\bf \hat{ n}}_{\alpha} =  \sin \left[ \alpha +\theta \right] \int\limits_0^{\infty} \hat{U}(k)     \exp \left[i k\frac{ \sin [\alpha+\theta]}{\cos \alpha} \left(x-x_1 \right)   -it\right] dk. \quad
      \label{u04na}
      \end{split}
 \end{equation}
 Substitution of $k \rightarrow  \gamma k$ in (\ref{u01na}), with $\gamma=\frac{\sin[ \alpha+\theta]}{\sin[ \alpha-\theta]}$, such that the exponential terms in (\ref{u01na}) and (\ref{u04na}) become identical, and inserting (\ref{u01na}) and (\ref{u04na}) into (\ref{impermeability}) gives
      \begin{equation}
 \begin{split}
 \Re \left[  \int\limits_0^{\infty} \left(\hat{U}(\gamma k)-\hat{U}(k)\right)  \exp \left[ i k  \frac{ \sin [\alpha+\theta]}{\cos \alpha} (x-x_1) - i t\right] dk \right] =0.
 \label{U_inclined_wall}
       \end{split}
 \end{equation}
Satisfying the free-slip boundary conditions at the four reflecting walls for all times $t$ thus imposes the spectral constraint
\begin{equation}
\hat{U}(\gamma k )= \hat{U}(k).
 \label{U(k)}
 \end{equation}
Solutions to this functional equation are non-unique, and can be expressed as $\hat{U}(k)=P(\log_{\gamma}(k))$ for arbitrary period-1 functions P  \citep[see][pp. 185 - 186]{BK16}.
For all these spectra (except $P=0$) the velocity expressions (\ref{U1}) and (\ref{U234}) are non-integrable for points on the inviscid wave attractor, $\zeta_n=0$, confirming the results by \cite{R01}. For all other points, the integrals in (\ref{U1}) and (\ref{U234}) are integrable only for discrete spectra $\hat{U}(k)$, i.e. the periodic function $P$ has to be a superposition of Dirac delta functions. The exact self-similar wave attractor solution by \citet{Ma09} in terms of countable infinite Fourier coefficients is an example having such a discrete spectrum $\hat{U}(k)$. The self-similar structure of wave attractors is reflected by the $\log_{\gamma}$-periodicity of the spectra. Next, we regularize the singularity by adding viscous attenuation, thereby also admitting continuous spectra. 

 \section{Internal shear layer dissipation}
\label{ivd}
Incorporating weak viscous attenuation in an asymptotic wave beam expression was first done by \cite{TS73}, and has been achieved using different procedures (see \S 6 of \cite{Vo03} for an overview). Here, we determine the effect of viscosity on the spectrum $\hat{U}(k)$ - an exponential attenuation factor - and incorporate it in the inviscid spectral decompositions for the velocity field, (\ref{U1}) and (\ref{U234}). We briefly demonstrate this analysis because of its similarity with the damping mechanisms caused by the rigid walls, presented in \S \ref{lateral} and \S \ref{refl}. \\
 For notational convenience, we drop the superscript $[n]$, and consider a wave attractor branch with velocity $U$ in the along-energy-propagation direction $\xi$, and phase propagation along $\zeta$. Upon incorporating continuity and buoyancy equations, one can write the governing equation for $U$ as
\begin{equation}
-\Delta U+N^2 \left(\sin^2 \theta \ U_{\zeta \zeta}+2 \cos \theta \sin \theta \ U_{\zeta \xi}+\cos^2 \theta \ U_{\xi \xi} \right) =-i \delta^2 \Delta^2 U.
\label{Umom}
\end{equation}
This equation is solved at $\mathcal{O}(\delta^0)$ by $U(\zeta)$ as defined in (\ref{U1}), provided the non-dimensional dispersion relation, $1= N^2 \sin^2 \theta$, holds. The velocity function $U$ is still an $\mathcal{O}(\delta^0)$-solution if we let the spectrum $\hat{U}(k)$ to be weakly dependent on the along-beam coordinate $\xi$, that is to say, if $\hat{U}_{\xi}\in \mathcal{O}(\delta)$. We assume $\hat{U}_{\xi} \in \mathcal{O}(\delta^2)\subset \mathcal{O}(\delta)$. Equation (\ref{Umom})  at $\mathcal{O}(\delta^2)$ then becomes
$$2 N^2 \sin \theta \cos \theta U_{\zeta \xi}=- i \delta^2 U_{\zeta \zeta \zeta \zeta}.$$
This is solved by 
\begin{equation}
U=\int_0^{\infty} \hat{U}(k,\xi) \exp \left[ i k\zeta-i t \right] dk, \quad  \quad \hat{U}(k,\xi)= \hat{U}(k) \exp \left[-\delta^2  \frac{ \tan \theta}{2} k^3 (\xi-\xi_0) \right],
\label{U(k,xi)}
\end{equation}
for arbitrary $\hat{U}(k)$, and where $\xi_0$ is the along-wave-attractor distance to the virtual localized source. Adding weak viscous attenuation to the 2D wave attractor velocity field is thus achieved by replacing
$\hat{U}(k) \rightarrow \hat{U}(k) \exp \left[-\delta^2  \frac{ \tan \theta}{2} k^3 (\xi-\xi_0) \right]$ in the velocity fields (\ref{U1}) and (\ref{U234}). \\
Note that the real (imaginary) part $U(\zeta,\xi)$ in (\ref{U(k,xi)}) is even (odd) in $\zeta$ around $\zeta=0$. This symmetry is preserved among reflections at horizontal or vertical boundaries, whereas reflections at inclined boundaries break it. All attractors include symmetry-breaking reflections, hence, their velocity fields cannot be symmetric around the inviscid attractor orbit, $\zeta=0$, when including viscous attenuation. Describing the wave attractor branches nevertheless by a viscous wave beam emitted from a virtual point source leads to a slight violation of the impermeability boundary condition. Physically, this means that the energy input into the fluid occurs through (non-uniform) oscillations of the wall, spatially at the scale of the cross-beam thickness.  For sufficiently long attractors, the asymmetry is small, and we proceed by neglecting it. \\
Incorporating the viscous attenuation in the impermeability constraint (\ref{U_inclined_wall}) at the reflection point $(x_1,z_1)$ results in the modified spectral constraint
\begin{equation}
\hat{U}(\gamma k )= \hat{U}(k) \exp \left[ -\delta^2  \frac{ \tan \theta}{2} k^3 \lambda \right] ,
 \label{U1(k)}
 \end{equation}
 where $\lambda=L_a/L_0$ is the non-dimensional length of the wave attractor. We consider $L_a \gg L_0$, such that the discussed asymmetry of the attractor is negligible. As a consequence, the attenuation rate $-\delta^2  \frac{ \tan \theta}{2} k^3 \lambda$ per attractor cycle can be orders of magnitude larger than $\mathcal{O}(\delta^2 k^3)$, namely if $\lambda \gtrsim \delta^{-1} \gg 1$ (note that by assumption, the most energetic wave number is non-dimensionalized to $2\pi$). \\ 
 The spectral constraint (\ref{U1(k)}) for the velocity field is equivalent to the constraint for the buoyancy gradient spectrum, $A(k)$, given by \cite{Ha08} upon correcting for a missing factor $1/2$ in their viscous attenuation rate, and a missing factor $\gamma^{-1}$ on the right-hand side of their recursive relation $A_n^2=\gamma^3 A_{n-1}^2$, where $A_{n}$ and $A_{n-1}$ are the buoyancy gradient spectra before and after the reflection from the slope, respectively. \\
 The constraint (\ref{U1(k)}) for $\hat{U}(k)$ now admits integrable finite-energy spectra:
 \begin{equation}
 \hat{U}(k)=P(\log_{\gamma}(k))\exp \left[ -\beta_1 k^3  \right], \qquad \mbox{ with } \ \  \beta_1= \frac{ \delta^2  \lambda \tan \theta}{2(\gamma^3-1)}
 \label{Uk1}
 \end{equation}
 for continuous period-$1$ functions $P$. The function $P(\log_{\gamma}(k))$ in the spectral solution (\ref{Uk1}) still reflects the geometric wave focusing, which projects the internal wave field distribution on any wave number interval $[k,\gamma k]$ onto $[\gamma k, \gamma^2 k]$, whereas the exponential term accounts for the energy dissipation upon traveling once around the wave attractor. If the energy input occurs within a low wave number interval, say $I_*=[k_*, \gamma k_*]$, with distribution $E(k)$, then $P(m)=\sqrt{E(\exp[\log(\gamma) m] )}$ defines $P(m)$ for all $m>log_{\gamma}(k_*)$ (by periodic continuation) and we take $P(m)=0$ for $m<\log_{\gamma}(k_*)$ (no energy at wave numbers smaller than $k_*$). If the energy input is spread over a wider interval than $I_*$, then one can split it into several intervals, define corresponding functions $P$ for each interval, and superimpose the resulting spectra.  For mathematical convenience, we take $P(m)$ to be periodic with period $1$ in the following.

 In the next section, we show that the dissipation at the lateral walls also adds an exponential attenuation factor to the spectral constraint (\ref{U1(k)}).
 
  \section{Dissipation at lateral walls}
 \label{lateral}
 In this section we extend the wave attractor velocity field to the lateral walls, $y=\pm l_y$, where we apply the no-slip boundary condition. Again, we do this for one (arbitrary) wave attractor branch with interior velocity field ${\bf u_0}={\bf \hat{\xi} } U$, and phase speed along ${\bf \hat{\zeta}}$. 
 Using the stretched coordinate  $\eta=\delta^{-1} y$, the momentum equations for $u_0$ and $w_0$ are given by
 \begin{equation} 
-i u_{0}= -p_{0_x}+u_{0_{\eta\eta}}, \qquad          i \cot^2 \theta \ w_{0}=-p_{0_z}+w_{0_{\eta\eta}}.
\label{2m}
\end{equation}
 In these two equations, the partial time derivatives have already been replaced by $-i$. It is the buoyancy, $ b_0=- i\sin^{-2} \theta \ w_0$, which adds to the time derivative of the vertical velocity component, $w_0$, producing the factor $-\cot^2 \theta$. Outside the boundary layers, the along-wave-beam velocity component $U$ is related to the pressure gradient in $\zeta$-direction by 
 \begin{equation}
 -i  \hat{\xi}_x U=-\hat{\zeta}_x p_{0_{\zeta}}  \quad \Rightarrow \quad        p_{0_{\zeta}}= -i  \cot \theta \ U ,
 \label{po}
 \end{equation}
 which solves the momentum equations in the unstretched coordinates at $\mathcal{O}(\delta^0)$, i.e. (\ref{2m}) without the diffusive terms. Here, $\hat{\xi}_x=\pm \cos\theta$ is the $x$-component of the unit vector ${\bf \hat{\xi}}$, and similarly $\hat{\zeta}_x=\pm \sin \theta$, the sign again depending on the branch. 
 Solving (\ref{2m}) with no-slip boundary conditions at the walls, $\eta=\pm \delta^{-1} l_y$, and interior velocity field ${\bf \hat{\xi}} U$ in the center plane, $\eta=0$, gives
\begin{equation}
u_0=   \hat{\xi}_x \left(1- \frac{\cosh[ i^{-\frac{1}{2}} \eta ]}{\cosh[ i^{-\frac{1}{2}} \delta^{-1} l_y ] }\right) U, \quad \ \ w_0=  \hat{\xi}_z \left(1- \frac{\cosh[   i^{\frac{1}{2}} \cot \theta \ \eta]}{\cosh[  i^{\frac{1}{2}} \cot \theta \ \delta^{-1} l_y ]} \right) U.
\label{uw}
\end{equation}
The presence of stratification (non-zero buoyancy) causes the factor-$\cot \theta$ difference in the thicknesses of the boundary layer, $\delta$ and $\delta \tan \theta$, for respectively horizontal and vertical velocity components, making $(u_0,0,w_0)$ divergent near the walls. This peculiar twist of the stratification on the boundary layer thickness was previously found by \cite{VC03} in their theoretical study on 3D internal wave generation by an inclined plane oscillating in the planar direction.  \\
Note that the $y$-momentum equation is satisfied at $\mathcal{O}(\delta)$ by choosing an appropriate pressure $p_2(\eta)$, which is $\mathcal{O}(\delta^2)$, thus negligible.
By the continuity equation at $\mathcal{O}(\delta^0)$ in stretched coordinate $\eta$,
$$ u_{0_x}+w_{0_z}=- v_{1_\eta}, $$
we get the $\mathcal{O}(\delta)$ transversal velocity component
\begin{equation}
v_1= \cos \theta \sin \theta \left(     i^{\frac{1}{2}} \frac{\sinh[ i^{-\frac{1}{2}} \eta ]}{\cosh[ i^{-\frac{1}{2}} \delta^{-1} l_y ] }              -            i^{-\frac{1}{2}} \tan \theta \frac{\sinh[   i^{\frac{1}{2}} \cot \theta \ \eta]}{\cosh[  i^{\frac{1}{2}} \cot \theta \ \delta^{-1} l_y ] }   \right)  U_{\zeta}+V(y),
\nonumber
\end{equation}
Here, $V(y)$ is an undetermined velocity component satisfying $V_{\eta}(y)\in\mathcal{O}(\delta)$, that is to say, slowly varying in the transversal $y$-direction. The impermeability boundary condition  ($v_1=0$) at both walls translates to
\begin{gather}
 V(\pm l_y)=\pm  \sigma  U_{\zeta}, \quad \mbox{with} \label{tildeV} \\
 \sigma= \cos \theta \sin \theta \left(         i^{-\frac{1}{2}} \tan \theta \tanh[   i^{\frac{1}{2}} \cot \theta \ \delta^{-1} l_y ]         -              i^{\frac{1}{2}} \tanh[ i^{-\frac{1}{2}} \delta^{-1} l_y ]   \right). \nonumber
\end{gather}
In the limit $\delta^{-1} l_y   \gg 1$, the expression simplifies to $\sigma=- i^{\frac{1}{2}} \sin \theta e^{i \theta }$. The transversal velocity component $V$ enters the continuity equation at $\mathcal{O}(\delta)$ in the unstretched coordinates: 
\begin{equation}
U_{\xi}+\delta V_y=0.
\label{U_cont}
\end{equation}
Since $U$ is $y$-independent, we get $V_{yy}=0$, hence
\begin{equation}
V= \frac{ \sigma y }{l_y  } U_{\zeta}.
\label{V}
\end{equation}
Thus, the transversal velocity $v$ decays linearly (hence slowly) towards the center plane, $y=0$, making the velocity field in the interior truly three-dimensional at $\mathcal{O}(\delta l_y^{-1})$.\\ 
The transversal divergence, 
$$V_y=\frac{ \sigma }{ l_y } U_{\zeta} = \frac{ i \sigma }{ l_y }   \int\limits_0^{\infty}  k \hat{U}(k)     \exp \left[i k\zeta    -it  \right] dk     ,$$ 
is balanced by $-\delta^{-1} U_{\xi}$, according to the continuity equation (\ref{U_cont}). This means that $U$ must be $\xi$-dependent at $\mathcal{O}(\delta)$. For the velocity expressions (\ref{U1}) and (\ref{U234}) of the wave attractor, this requires the spectrum $\hat{U}(k)$ to be replaced by $\hat{U}(k) \exp \left[ -i \delta l_y^{-1} \sigma k \xi\right]$. Consequently, the velocity $U$ decays in the along-wave-beam direction, $\xi$, with $\exp[-\delta l_y^{-1} \sigma_0 k \xi ]$, where $\sigma_0=\Re[i \sigma]>0$ for $\theta\in(0,\pi/2)$. The imaginary part of $ i l_y^{-1} \sigma$, which takes both positive and negative values for $\theta\in(0,\pi/2)$, describes a slight change in tilt in phase propagation direction, that changes from $\zeta$ to $\zeta- \delta l_y^{-1} \Re[\sigma] \xi$. \\
Adding the damping by the lateral walls to the constraint for the 2D viscous wave attractor spectrum, (\ref{U1(k)}), gives
\begin{equation}
\hat{U}(\gamma k )= \hat{U}(k) \exp \left[ \left( -\delta^2  \frac{ \tan \theta}{2} k^3 -i \delta \frac{\sigma}{l_y }k\right) \lambda \right] .
 \label{U2(k)}
 \end{equation}
This extended equilibrium wave attractor spectrum constraint is solved by
\begin{equation}
\hat{U}(k)=P(\log_{\gamma}(k)) \exp \left[ -\beta_1 k^3 - \beta_2 k \right]   , \qquad \mbox{ with } \ \ \beta_2=\frac{i \delta \lambda   \sigma}{l_y (\gamma-1)} ,
 \label{Uk2}
 \end{equation}
for all period-$1$ functions $P$. 

  \section{Dissipation at reflecting walls}
 \label{refl}
 
\begin{figure}
\begin{center}
\begin{tikzpicture}

\fill[fill=gray!20] (-1.45, -2.4) -- (0.4, -0.6) -- (-0.4, 0.6) -- (-3.4, -2.4); 
\fill[fill=gray!20] (-3.6  ,3.3) -- (0.4, -0.6) -- (-0.4, 0.6) -- (-2.99 ,3.3);  

\draw (-2.2,3.3) --  (2.4*2/3,-2.4) node[pos=0.87,above,sloped] {inclined wall}; 

\draw[dashed,-] (-2.4,-2.4) -- (0,0) node[pos=0.3,above,sloped] {incident beam} ; 
\draw[dashed,-]   (0.15,-0.15) --(-3.3,3.3)  node[pos=0.65,below,sloped] {reflected beam} ;

\draw[thick,->] (1,1.5) -- (1.5, 1.5) node[anchor=west] {x}; 
\draw[thick,->] (1,1.5) -- (1,2) node[anchor=south] {z};

\draw[thick,->] (-4/3 ,   2) -- (-4/3  +0.53,2    +0.53*2/3) node[anchor=south west] {z'}; 
\draw[thick,->] (-4/3.03  ,  2) -- (-4/3.03  +0.53*2/3,2   -0.53) node[anchor= south west] {x'};

\draw[ultra thick,->] (-1.75,-1.75) -- (-1.1,-1.1) node[pos=0.8, below] {${\bf u}^{[4]}_0$}; 

\draw[ultra thick, dashed, ->] (-1.08,1.13) -- (-3.08,3.13) node[pos=0.8,above] {$\ \ {\bf u}^{[1]}_0$}; 
\draw[ultra thick,->] (-1.1,1.1) -- (-2.7,2.7) node[pos=0.57, above] {$\quad \ \ {\bf u}^{[1]}\ \  $}; 

\draw[ultra thick,->] (0,0) -- (0.4,-0.4) node[anchor=west] {$\delta {\bf u}^{[1]}_1 $};  			    

\draw[ultra thick,->]  (0.66*0.5-0.1   -0.1   ,-0.5-0.1   +0.1   ) --(-0.66*0.5-0.15     -0.1 ,0.5-0.1   +0.15    )  ;
\draw[ultra thick,->]  (0.66*0.25-0.25     -0.125     ,-0.25-0.25    +0.125)   --   (-0.66*0.25-0.25         -0.15 ,0.25-0.25        +0.15)    node[pos=0.15,below] {$\tilde{u}'_0$};
\draw[ultra thick,->] (0.66*0.1-0.4    -0.15   ,-0.1-0.4    +0.15) -- (-0.66*0.1-0.4     -0.15  ,0.1-0.4    +0.15 )  ;

\draw[ultra thick,->] (0.2, 0.15) -- (0,0) node[pos=0.3, above] {$\delta  \tilde{w}'_1$};

\draw[very thin]  (2.2*2/3  -0.4,-2.2   +0.6) arc (-240:-180:0.7) ; 
\draw[very thin] (2.2*2/3  -0.75,-2.2 ) --    (2.2*2/3 ,-2.2) node[pos=0.45, above] {$\alpha$};

\draw[very thin]  (-1.7, -1.7) arc (45:0:0.7) ;  
\draw[very thin](-2.2, -2.2)--    (-2.2+0.7, -2.2 )node[pos=0.65, above] {$\theta$};

\end{tikzpicture}
\caption{This sketch illustrates the velocity components involved in the reflection of the wave attractor (dashed line, wave motion confined to shaded area) at the inclined wall (solid line), with arrows pointing in the energy propagation directions. The velocity field of the reflected beam, ${\bf u}^{[1]}={\bf u}^{[1]}_0+\delta {\bf u}^{[1]}_1+\mathcal{O}(\delta^2)$, is - due to focusing - larger in amplitude than the incident beam velocity, ${\bf u}^{[4]}_0$, but weaker than the free-slip reflected velocity field, ${\bf u}_0^{[1]}$ (dashed arrow). The velocity component $\tilde{u}'_0$, pointing along the wall, annihilates the free-slip velocity field ${\bf u}^{[1]}_0+{\bf u}^{[4]}_0$ at the wall, $z'=0$, and decays exponentially towards the interior. By mass conservation, it generates the velocity component $\tilde{w}'_1$ at $\mathcal{O}(\delta)$, normal to the wall, which itself is canceled by ${\bf \hat{z}' } {\bf u}^{[1]}_1$ at $z'=0$. Contrary to $\tilde{u}_0$ and $\tilde{w}_1$, the component ${\bf u}_1^{[1]}$ does not decay towards the interior; it is the correction on the reflected beam due to damping by the no-slip reflection. }
\label{sketch1}
\end{center}
\end{figure}
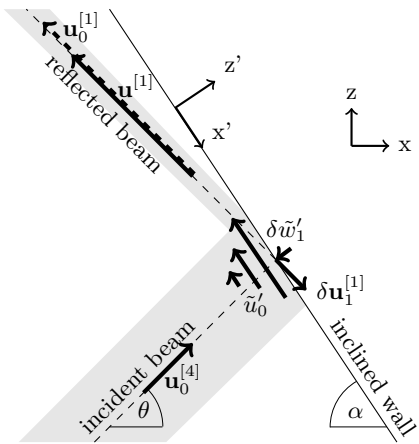

 No-slip reflection of 2D monochromatic internal waves from a wall has been analyzed theoretically 
 for wave beams by \cite{KC95b, KC95a}. Whereas dissipation due to internal shear is included in the analysis by \cite{KC95b, KC95a}, they do not account for the energy loss in the viscous boundary layer, which also weakens the reflected wave beam. We are interested in precisely this energy loss at the reflecting wall, such that we can tell when it is negligible. \\
 To begin with, we consider the inviscid free-slip velocity field at the inclined wall, $z=(l_x-x) \tan \alpha$, as this is the most general prescription of a planar reflecting boundary.  Expressed in the rotated and shifted coordinate system of the inclined wall,  
 $$
 \begin{bmatrix}
         x'\\
       z'
     \end{bmatrix}=      \begin{bmatrix}
       \cos \alpha \  - \sin \alpha   \\
      \sin \alpha  \ \ \  \cos \alpha
     \end{bmatrix} \cdot  \begin{bmatrix}
       x-x_1  \\
       z-z_1
     \end{bmatrix},   $$
 with $z'$ normal to the wall (see sketch in  Fig. \ref{sketch1}), the inviscid free-slip velocity field at the inclined wall, $z'=0$, is given by
 \begin{gather}
  {\bf u}_0^{[1]}+ {\bf u}_0^{[4]}= {\bf \hat{x}'} \tilde{U}(x'),    \label{free_slip_u0}  \qquad \quad
    \tilde{U}(x')  = \frac{- \sin 2 \theta }{ \sin [\alpha- \theta] }\int\limits_0^{\infty} \hat{U}(k) \exp \left[ i k  \sin [\alpha+ \theta] x' - i t \right] dk,   \nonumber
 \end{gather}
 where we have used the velocity expressions from \S \ref{interior} and the  inviscid spectral constraint (\ref{U(k)}).

 The task is now to find a quasi-2D correction velocity field, ${\bf \tilde{u}}=[\tilde{u},0,\tilde{w}]$, such that it annihilates the free-slip velocity (\ref{free_slip_u0}) at the inclined wall, $z'=0$, and decays exponentially towards the interior. 
 Using the stretched coordinate $Z= \delta^{-1} z'$, the $x'$-momentum equation at $\mathcal{O}(\delta^0)$, governing the velocity component $\tilde{u}_0'=\tilde{u}_0 \cos \alpha- \tilde{w}_0 \sin \alpha$ in the direction along the inclined wall, becomes
 \begin{equation}
 i \left(\frac{\sin^2 \alpha }{\sin^{2} \theta}-1 \right) \tilde{u}_0'  = \tilde{u}_{0_{Z Z}}'   .
  \label{m2A}
 \end{equation}

 \noindent As previously in (\ref{2m}), we have replaced the partial time derivatives with $-i$, and used $\tilde{b}_0= i \frac{  \sin \alpha }{\sin^{2}\theta} \tilde{u}_0' $. 
 The pressure gradient is absent because the pressure is not modified by the no-slip boundary.   
 Solving (\ref{m2A}) for ${\bf \tilde{u}}_0= {\bf \hat{x}'} \tilde{u}_0'$ such that it annihilates (\ref{free_slip_u0}) at $z'=Z=0$ and vanishes in the interior, $Z '\rightarrow -\infty$, gives
 \begin{equation}
 {\bf \tilde{u}}_0 =   - {\bf \hat{x}'}  \tilde{U}(x') \exp \left[ i^{\frac{1}{2}} \mu  Z \right] \qquad \mbox{ with } \ \mu =\sqrt{\frac{\sin^2 \alpha }{\sin^{2} \theta}-1}. 
 \label{u_alpha}
 \end{equation}
  By the continuity equation at $\mathcal{O}(\delta^0)$ in stretched coordinate $Z$,
 $$  \tilde{u}_{0_{x'}}' + \tilde{w}_{1_{Z}}'=0,$$
 with $\tilde{w}_{1}'$ the $\mathcal{O}(\delta)$-velocity component normal to the wall,  we get
\begin{equation}
\tilde{w}_1'=  i^{-\frac{1}{2}} \mu^{-1} \tilde{U}_{x'}(x') \exp \left[ i^{\frac{1}{2}} \mu  Z \right]  + F(x', z').
\label{tilde_v1}
\end{equation}
Here, $F$ is an undetermined velocity component, with spatial variations of $\mathcal{O}(1)$, similar to $V$ in the previous section. Previously, we were able to find a linear function in $y$ for $V$, such that the impermeability boundary conditions at opposite lateral walls are satisfied. This procedure fails here, and we must take $F=0$. As a consequence, $\tilde{w}_1'$ describes an apparent flow through the inclined wall, $z'=Z=0$.  This apparent flow through the wall,
\begin{equation}
\tilde{w}_1'(x',Z=0)=   - i^{\frac{1}{2}} \frac{\gamma \sin 2 \theta} {\mu} \int\limits_0^{\infty} k \hat{U}(k) \exp \left[ i k  \sin [\alpha+ \theta] x' - i t \right] dk ,
\label{v1}
\end{equation}
 can be balanced by absorbing some $\mathcal{O}(\delta)$-fraction of the incident wave beam (see also illustration in Fig. \ref{sketch1}). Consequently, the viscously reflected beam with velocity field ${\bf u}^{[1]}={\bf u}_0^{[1]}+\delta {\bf u}_1^{[1]}+\mathcal{O}(\delta^2)$ is weaker than the inviscid velocity field, ${\bf u}_0^{[1]}$. We write 
 \begin{equation}
 {\bf u}^{[1]}=\hat{\xi}_1 \int\limits_0^{\infty} \hat{U}(k) \exp[-\delta R_{\alpha} k]\exp \left[ i k \zeta_1-i t \right] dk, 
 \nonumber
 \end{equation}
 such that
  \begin{equation}
{\bf u}_1^{[1]}=-\hat{\xi}_1 R_{\alpha} \int\limits_0^{\infty} k \hat{U}(k)\exp \left[ i k \zeta_1-i t \right] dk, 
 \label{u11}
 \end{equation}
 where $\delta  \Re[R_{\alpha}]>0$ is the dissipation rate (per wave number) due to the reflection. 
 
  \begin{figure}
\begin{center}
  \hspace*{-0cm}
\includegraphics[type=pdf,ext=.pdf,read=.pdf, width=0.5\textwidth]{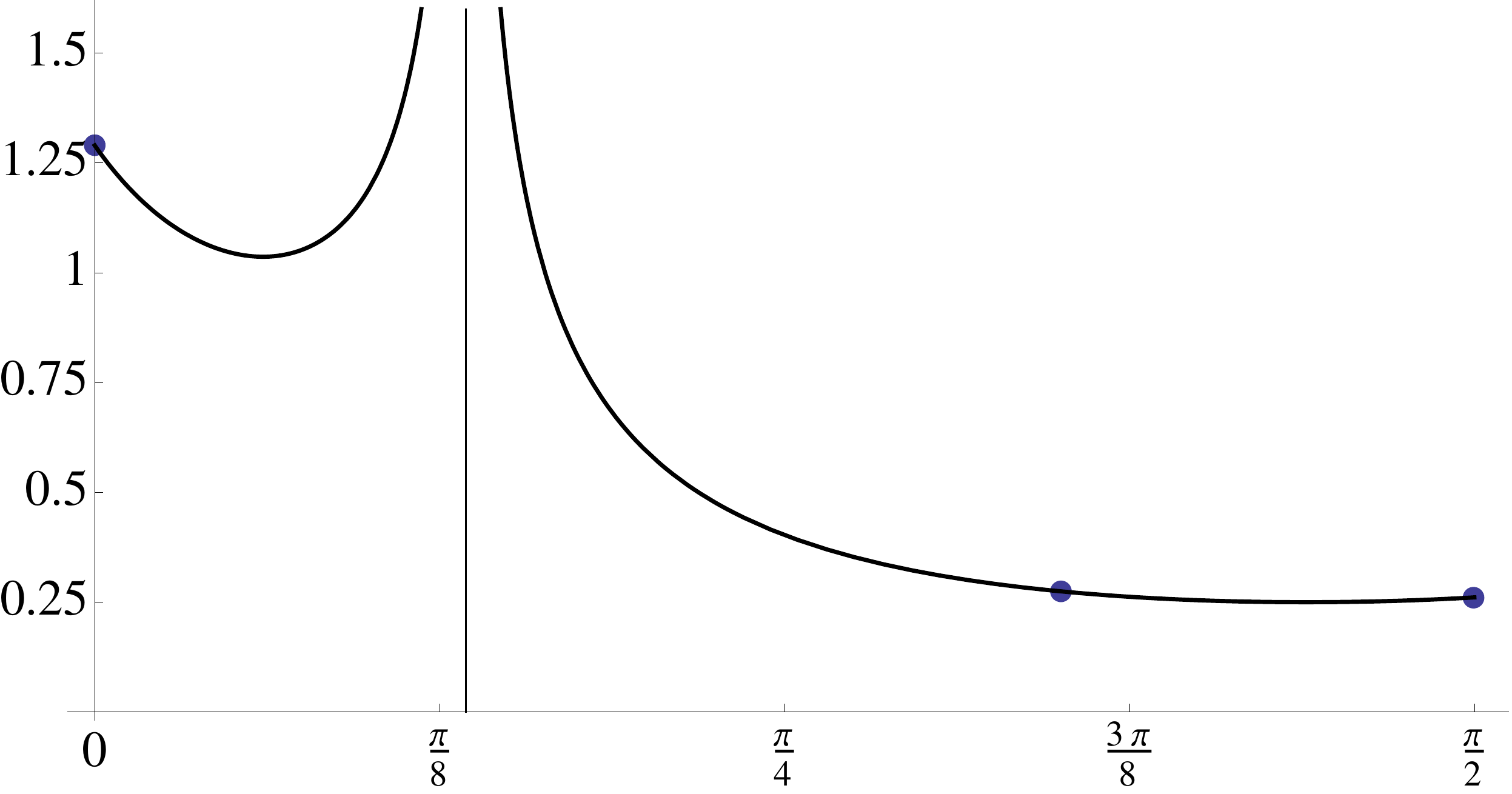}
\put(-220,92){$\Re[R_{\varphi}]$}
\put(2,8){$\varphi$}
\put(-178,83){$\Re[R_0]$}
\put(-58,28){$\Re[R_{\alpha}]$}
\put(-4,27){$\Re[R_{\pi/2}]$}
\caption{This figure presents the $\mathcal{O}(\delta)$-dissipation rate, $\Re[R_{\varphi}]$, as a function of the angle $\varphi$ of the reflecting wall with respect to the vertical, for $\theta=0.42$ rad. As to be expected, $R_{\varphi}$ blows up at the critical reflection angle, $\varphi=\theta=0.42$ (vertical asymptote). The total dissipation by no-slip reflections of a wave attractor such as in Fig. \ref{f1}a is the sum of dissipation rates at the bottom ($\Re[R_0]$), the inclined wall ($\Re[R_{\alpha}], $ here $ \alpha=1.1$) and the vertical wall ($\Re[R_{\pi/2}]$).  }
\label{R_fig}
\end{center}
\end{figure}
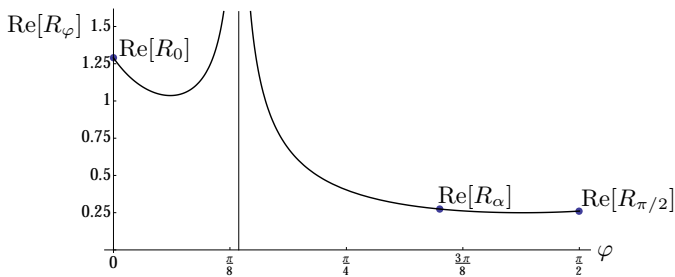

\noindent The complex-valued reflection dissipation rate $R_{\alpha}$ is determined by the impermeability condition at $\mathcal{O}(\delta)$:
\begin{equation}
\tilde{w}_1'+{\bf u}_1^{[1]} \cdot {\bf \hat{z}'}=0 \qquad \mbox{ at } \quad z'=0.
 \label{impermeability_delta1}
 \end{equation}
Substituting (\ref{v1}) and (\ref{u11}) into (\ref{impermeability_delta1}) and noting that on the inclined wall, $z'=0$, we have $ \zeta_1=\sin[\alpha-\theta] x'$, gives
\begin{equation}
R_{\alpha}=i^\frac{1}{2} \frac{ \sin 2 \theta}{ \mu \sin[\alpha-\theta] } .
\label{R_sol}
\end{equation}
We can readily use expression (\ref{R_sol}) to determine the dissipation rates due to the reflections at respectively the flat bottom ($\alpha \rightarrow 0$) and the vertical wall ($\alpha  \rightarrow \pi/2$):
$$R_0=i^{-\frac{1}{2}} 2 \cos \theta \qquad 			\mbox{and} 			\qquad R_{\pi/2}=i^{\frac{1}{2}}  2   \sin \theta \tan \theta = i \tan^2 \theta R_0.$$ 
The dissipation rate (real part of Eq. (\ref{R_sol})) as a function of the angle of the reflecting boundary is shown in Fig. \ref{R_fig}. In laboratory and numerical set-ups, the surface of the fluid, $z= h $, is typically free, so the most appropriate constraint on this boundary is free-slip (because vertical variations are negligibly small), i.e. no dissipation by reflection. The full viscous 3D equilibrium wave attractor spectrum must thus satisfy 
\begin{equation}
\hat{U}(\gamma k )= \hat{U}(k) \exp \left[ -\delta^2   \left( \lambda/2 \tan \theta \right) k^3 -\delta \left( i \sigma   l_y^{-1} \lambda + R_{\alpha} +R_0+R_{\pi/2} \right) k  \right] .
 \label{U3(k)}
 \end{equation}
Solutions to this spectral constraint are given by
\begin{equation}
\hat{U}(k)=P(\log_{\gamma}(k)) \exp  \left[ - \beta_1 k^3 - \beta_2 k -\beta_3 k \right]   , \quad \mbox{ with } \ \ \beta_3=\delta \frac{R_{\alpha}+R_0+R_{\pi/2} }{\gamma-1} ,
 \label{Uk3}
 \end{equation}
 for arbitrary period-$1$ functions $P$. If not stated else wise we always consider $P=$ constant in the following.

\section{Comparison with laboratory experiments and 3D simulations}
\label{comparison}
We validate our theoretical results by comparing it with experimental spectral results by \cite{Ha08} and  \cite{Brouzet2016} in \S \ref{c1} and \S \ref{c2} respectively.  \S \ref{c2} also includes a comparison with fully 3D numerical simulations, replicating one of the experiments by \cite{Brouzet2016}.

\subsection{Comparison with laboratory experiment by \cite{Ha08}}
\label{c1}
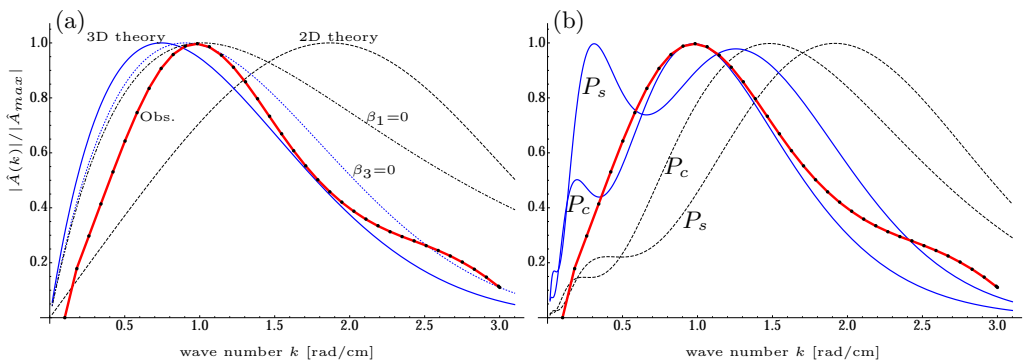
\begin{figure}
\begin{center}
\begin{minipage}{.485\textwidth}
  \hspace*{-0.0cm}
\includegraphics[type=pdf,ext=.pdf,read=.pdf, width=1\textwidth]{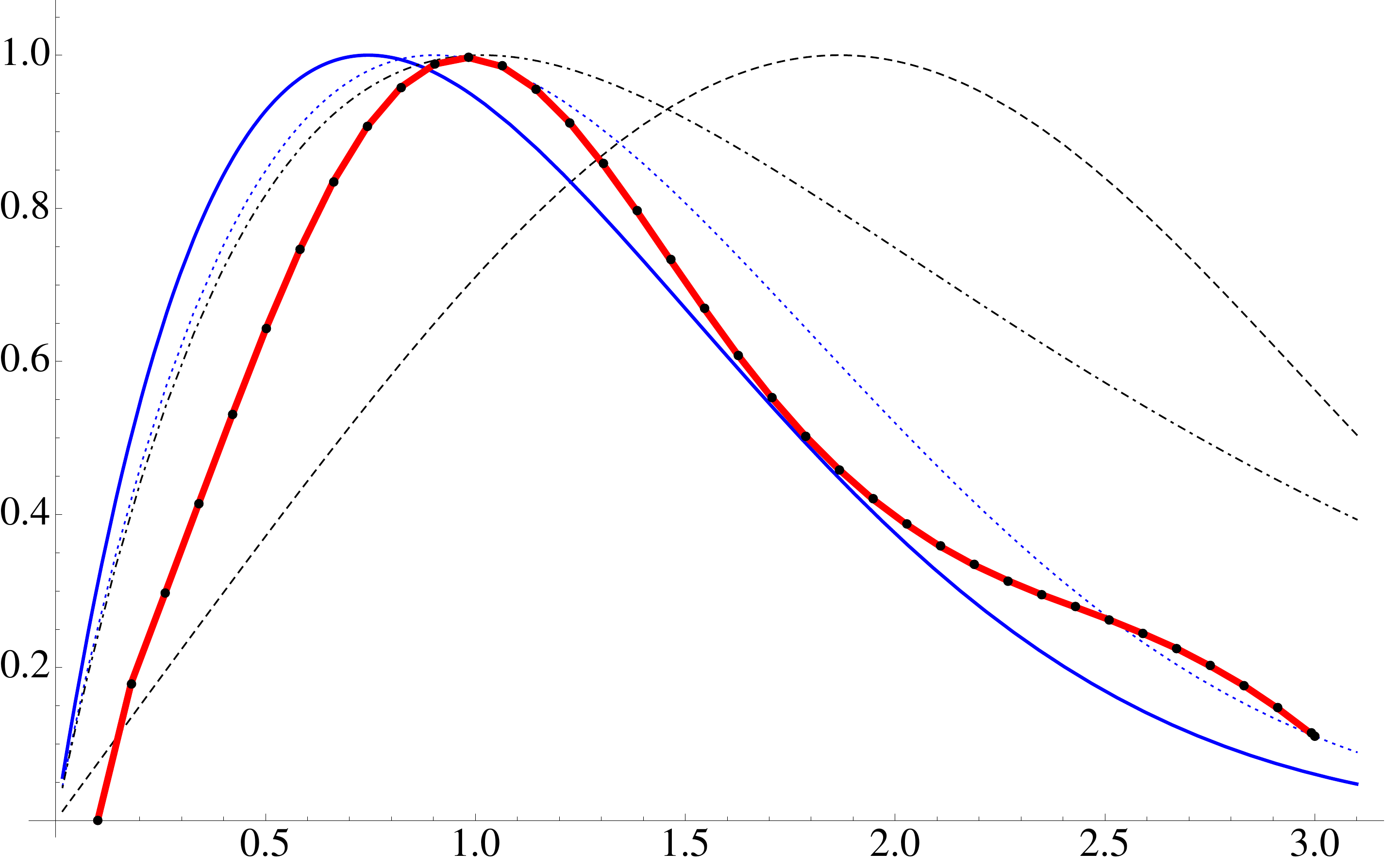}
   \put(-195,50){\rotatebox{90}{ \tiny $ | \hat{A}(k) | / | \hat{A}_{max} |  $ } }
    \put(-130,-8){\tiny wave number $k $ [rad/cm] }
    \put(-165,110){\tiny 3D theory}
    \put(-85,110){\tiny 2D theory}
    \put(-145,80){\tiny Obs.}
    \put(-60,80){\tiny $\beta_1$=$0$}
    \put(-65,60){\tiny $\beta_3$=$0$}
    \put(-180,115){ (a)}
    \end{minipage}
\begin{minipage}{.485\textwidth}
\includegraphics[type=pdf,ext=.pdf,read=.pdf, width=1\textwidth]{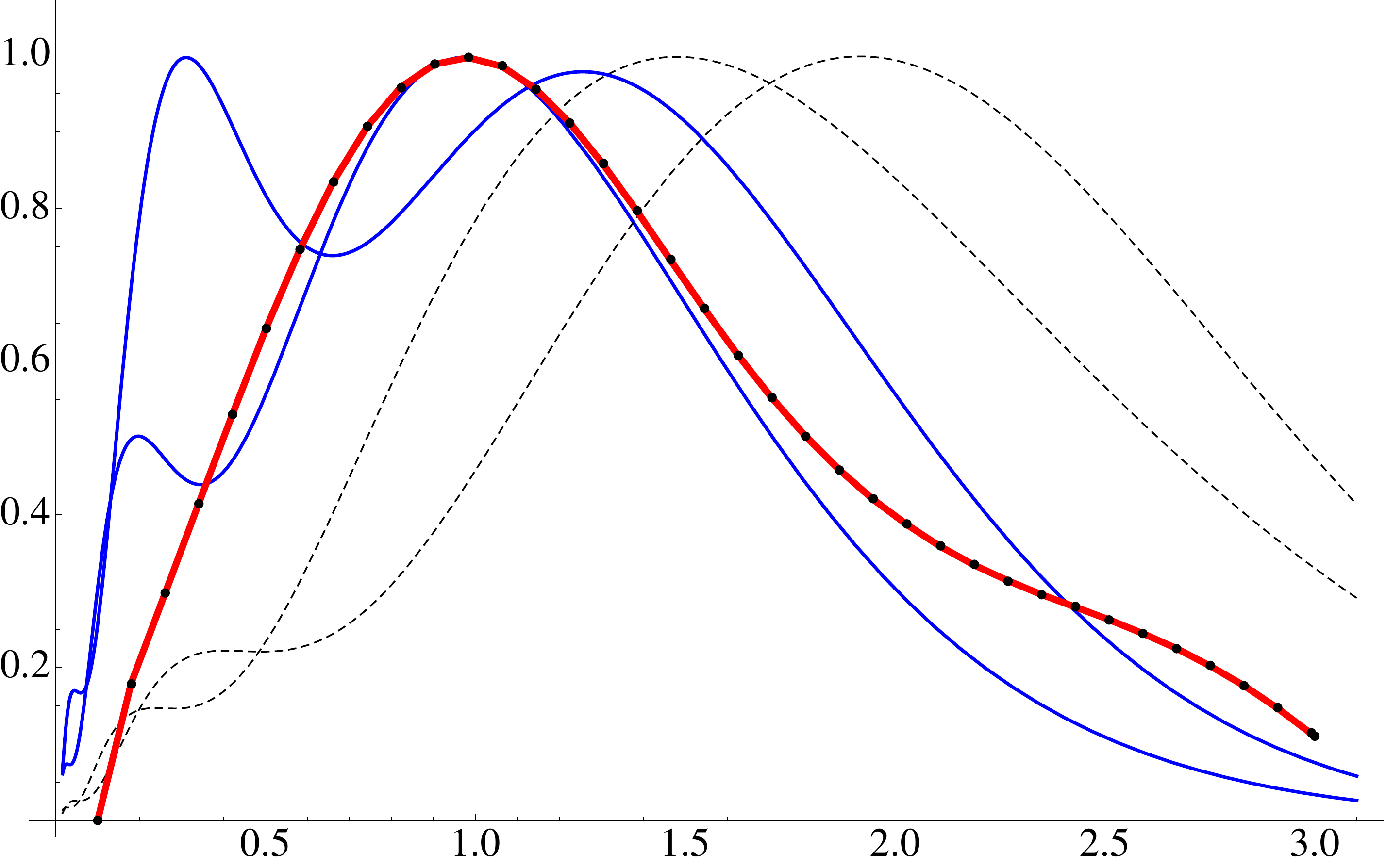}
  \put(-130,-8){ \tiny wave number $k $ [rad/cm]}
  \put(-166,90){$P_s$}
   \put(-172,46){$P_c$}
   \put(-135,60){$P_c$}
   \put(-128,40){$P_s$}
    \put(-180,115){ (b)}
\end{minipage}
\caption{Normalized modulus of the buoyancy gradient spectra, $\hat{A}(k)\propto k \hat{U}(k)$. The red curves with black dots reproduce the observed spectrum by \cite{Ha08}. Theoretical spectra are presented in plot (a) with $P=$ constant and in plot (b) with $P_s(k)=3+\sin(2\pi k)$, $P_c(k)=3+\cos(2\pi k)$: 3D spectrum (Eq. (\ref{Uk3}), thick blue) and 2D spectrum (Eq. (\ref{Uk1}), black dashed). Plot (a) also shows the spectra excluding dissipation upon reflection (Eq. (\ref{Uk2}), i.e. $\beta_3=0$, dotted blue) and excluding internal shear dissipation (Eq. (\ref{Uk3}) with $\beta_1=0$, black dashed-dotted).}
\label{f4}
\end{center}
\end{figure}

\cite{Ha08} studied the equilibrium spectrum of  internal wave attractors in the classical trapezoidal set-up, both in the laboratory and with a simple model. The parameter values relevant for the comparison with our theory are listed in Table \ref{t2}. 
Using synthetic schlieren techniques, they directly measured the buoyancy gradient field, $[b_x,b_z]$. Spatial variations of this buoyancy gradient field for each wave attractor branch are primarily in the corresponding phase propagation directions, $\zeta$. Fig. \ref{f4} reproduces the normalized modulus of the observed spectrum $\hat{A}(k)=\frac{-i }{\sin\theta} k \hat{U}(k)$ of the buoyancy gradient, $b_{\zeta}$, pointing in the phase propagation direction of the first wave attractor branch, along transect $S_1$ as shown in figure 3.1(a) in \cite{Ha08}. For comparison, Fig. \ref{f4} shows our theoretical 3D wave attractor spectrum (thick blue solid line) for different period-$1$ functions $P$ in Eq. (\ref{Uk3}). Additionally, we present the theoretical 2D spectrum including internal shear dissipation only (Eq. (\ref{Uk1}), dashed line in Fig. \ref{f4}), which corresponds to the 2D theoretical spectrum by \cite{Ha08}, their Eq. 3.7, for $P=$ constant and upon correcting mathematical mistakes in their analysis, mentioned in \S \ref{ivd}. Note that \cite{Ha08} seemingly achieved a good fit in their figure 6 because they changed their input wave number, $k_{in}=2\pi/H$ (in their notation $k_0$), while keeping the same $k_{in}$ fixed in their Eq. 3.7. Correct application of their theory reveals that their theoretical spectrum does not depend on their input wave number, $k_{in}$, and that the theoretical 2D spectrum predicts the attractor wave length to be a factor 2 smaller than observed.  The mismatch between the 2D spectra and observation supports our striking and unexpected conclusion that dissipation at the rigid walls must be substantial. \\
To illustrate the importance of the different dissipation mechanisms, we also present in Fig. \ref{f4} the spectra excluding dissipation upon reflection (Eq. (\ref{Uk2}), dotted blue line) and excluding internal shear dissipation (Eq. (\ref{Uk3}) with $\beta_1=0$, black dashed-dotted line) for $P =$ constant. \\
Three conclusions can be directly inferred from the comparison in Fig. \ref{f4}.

i) The full 3D wave attractor spectrum fits the observed spectrum reasonably well for the choices $P=$ constant and $P(k)=P_c(k)=3+\cos (2\pi k)$. The contact surface of the wave attractor with the tank boundaries (shaded surfaces in Fig. \ref{f1}a) consists primarily ($\sim$ 73\%) of those at the lateral walls. It thus comes as no surprise that in this particular laboratory set-up, with $\beta_3\approx 0.28 \beta_2$, neglecting dissipation at the reflecting walls still results in good fits with the observation (see also relatively small difference between solid and dotted blue lines in Fig. \ref{f4}a). Hence, dissipation occurs primarily in the internal shear layers and in the lateral boundary layers, and secondarily also at the reflecting rigid boundaries.

ii) Neglecting internal shear dissipation (Eq. (\ref{Uk3}) with $\beta_1=0$, dashed-dotted line) leads to a spectrum whose peak coincides with the observation. However, at large wave numbers, this spectrum diverges from the observation. This indicates that the neglected internal shear dissipation, which is cubic in wave number $k$, is the dominant dissipation mechanism at high wave numbers in the laboratory experiment.

\begin{table}
\centering
\label{my-label}
\begin{tabular}{llll}
\BV frequency			 					& $N_0$													& $3$		& rad/s 		\\
Angle of wave beam with respect to horizontal		& $\theta=\arcsin[\omega_0/N_0]$								& $0.42$		& rad		\\
Angle of sloping wall with respect to horizontal		& $\alpha$ 												& $1.10$		& rad 		\\
Width of the tank							& $W=2 l_y L_0 $											& $10.1$ 		& cm			\\
Tank length at bottom						& $L =2 l_x L_0$											& $45.3$		& cm			\\
Water column height							& $H=h L_0$												& $19.0$		& cm			\\
Wave attractor length						& $L_a=\lambda L_0 $										& $85.0$		& cm 		\\
\end{tabular}
\caption{Parameter values of the laboratory experiment by \cite{Ha08}}
\label{t2}
\end{table}

iii) The discrepancy between full 3D spectrum for $P_c(k)=3+\cos (2\pi k)$ and $P_s(k)=3+\sin (2\pi k)$ shows that the shape of the theoretical spectrum depends strongly on this period-$1$ function, $P$. As discussed in \S \ref{ivd}, the precise nature of the $P$ is set by the spatial structure of the energy input, i.e. by the geometry of the tank used in the experiment by \cite{Ha08}. This means that the energy input strongly influences the spatial structure of the equilibrium wave attractor, and upscaling of a laboratory set-up generally does not leave the wave attractor invariant. Despite the sensitivity on $P$, we can only achieve reasonable fits between theory and observations if we include dissipation at the rigid boundaries.

\begin{figure}
\begin{center}
         
\includegraphics[type=pdf,ext=.pdf,read=.pdf, width=1\textwidth]{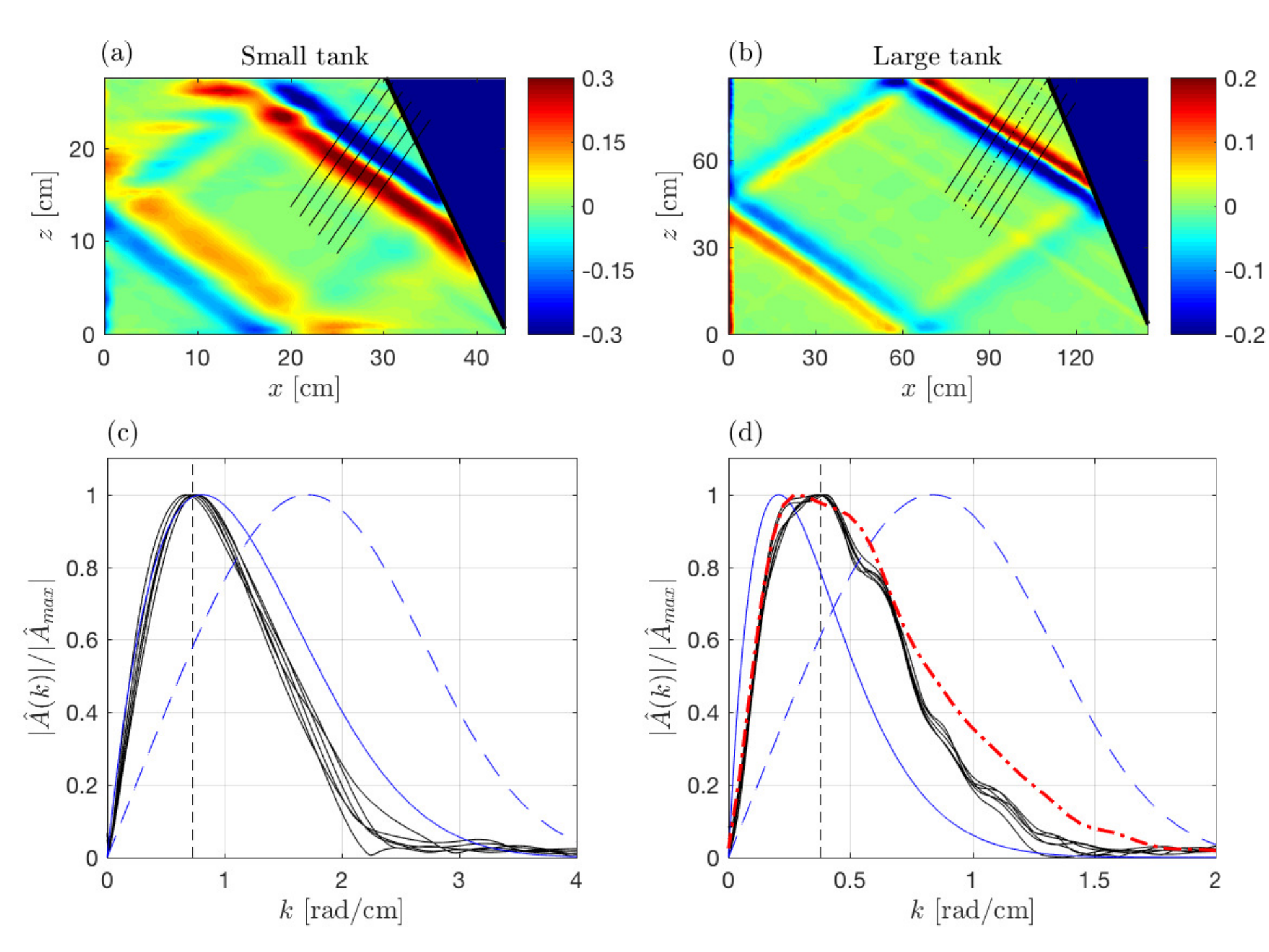}

\caption{ Experimental results by \cite{Brouzet2016} for steady state wave attractors in the small and large tank set-ups. Top: snapshots of the buoyancy gradient, $b_{\zeta}$, in the phase propagation direction $\zeta$ of first branch, derived from observed field $[b_x,b_z]$ after Hilbert-filtering at $\omega_0$. Bottom: Normalized modulus of experimental buoyancy gradient spectra, $|\hat{A}| / |\hat{A}_{max} |$, (black lines) along the depicted transects of the first attractor branch in top panels. For comparison, corresponding 3D theoretical spectra (solid blue) and 2D spectra (dashed blue). The red dot-dashed curve in (d) shows the numerical spectrum taken along dot-dashed transect in (b) after $\omega_0$-Hilbert filtering simulated steady state time-series of large-tank set-up.   }
\label{f1_exp}
\end{center}
\end{figure}

\subsection{Comparison with laboratory experiments by \cite{Brouzet2016} and 3D simulation}
\label{c2}

\begin{table}
\centering
\label{my-label}
\begin{tabular}{lllll}
~					 					&         	~												& Small tank     & Large tank		& 	~	      \\
\BV frequency			 					& $N_0$													& $1.37$ 		& $0.867$ 		& rad/s 		\\
Angle of wave beam w.r.t. horizontal				& $\theta=\arcsin[\omega_0/N_0]$								& $0.61$ 		& $0.58$  			& rad		\\
Angle of sloping wall w.r.t. horizontal				& $\alpha$ 												& $1.13$ 		& $1.18$	 		& rad 		\\
Width of the tank							& $W=2 l_y L_0 $											& $17.0$ 		& $17.4$ 			& cm			\\
Water column height							& $H=h  L_0$ 												& $29.5$		& $92.0$ 			& cm			\\
Wave attractor length						& $L_a=\lambda L_0 $										& $103.3$		& $337.8$		 	& cm 		\\
Wave maker amplitude						& $a$													& $2.5$ 		& $1.5$ 			& mm		\\
\end{tabular}
\caption{Parameter values of the laboratory experiments by \cite{Brouzet2016}. }
\label{t1}
\end{table}

\begin{figure}
                \centering
                \includegraphics[type=png,ext=.png,read=.png, width=0.970\textwidth]{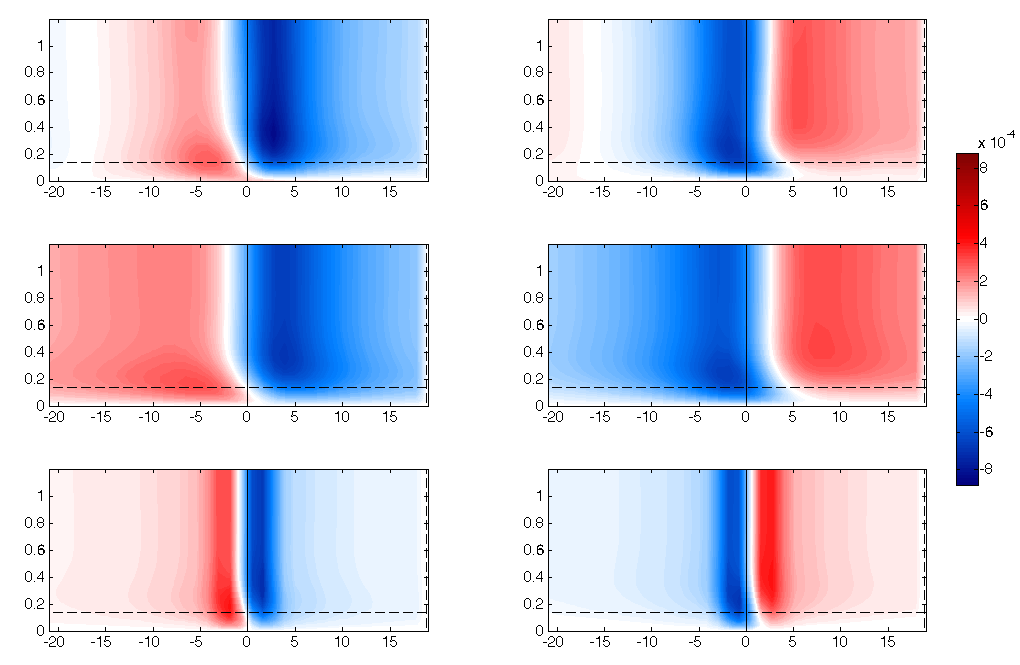}
                \put(-145,240){Numerical, $t=1004$ s}                
                \put(-330,240){Numerical, $t=1000$ s}   
                \put(-330,156){Theory, 3D spectrum}        
                \put(-140,156){Theory, 3D spectrum}        
                \put(-330, 74){Theory, 2D spectrum}        
                \put(-140, 74){Theory, 2D spectrum}        
                \put(-375,195){ \tiny \rotatebox{90}{ $y$ [cm] } }
                \put(-375,110){ \tiny \rotatebox{90}{ $y$ [cm] } }
                \put(-375,30){ \tiny \rotatebox{90}{ $y$ [cm] } }
                \put(-291,2){ \tiny $\zeta$ [cm]}
                \put(-108,2){ \tiny $\zeta$ [cm]}
                \caption{Top: Snapshots of simulated density perturbation $[g/cm^3]$ ($\propto b$ since salt diffusion is negligible small) at times $t=1000$ s and $t=1004$ s (0.3 wave periods apart) in an inclined phase propagation plane, $\xi=$ constant, intersecting the first attractor branch. Middle: Theoretical buoyancy, $b$, at corresponding phases for 3D spectrum. Bottom: Same theoretical $b$, but with 2D spectrum. The (buoyancy) boundary layer widths, $d_0 \tan \theta$ and $d_0 \mu$ at respectively lateral wall ($y=0$) and inclined wall ($\zeta=19$ cm), are indicated by the dashed lines; the solid line shows the center of the wave attractor.  }
\label{comparison_with_buoyancy}
\end{figure}

\cite{Brouzet2016} performed laboratory experiments on wave attractors in two trapezoidal tanks with almost identical lateral widths ($2W$), but with differences in height ($H$) and length ($L$) of approximately a factor $3$ (see Table \ref{t1} for parameter values). Here, we briefly describe the experiments for a comparison with our theory.\\
In both experimental set-ups, the internal waves are generated by a sinusoidally shaped wave maker \citep{Go07} situated on the left side of the tank, with the vertical wave length corresponding to half the height of the water column, so $k_{in}=\frac{\pi}{H \cos \theta}$. Previous experiments, also reported in \cite{Br16a, Br16,Br17}, show that triadic resonance instabilities arise if the wave maker amplitude, $a$, exceeds a critical values in the range $2.5 -3 $ mm,  dependent on the position of the attractor. Both experiments presented here are stable, and a steady state is reached after a spin-up of roughly $20$ wave periods. According to our theory, both laboratory set-ups fall into a regime where both internal shear and rigid-wall dissipation are significant. Hence, we expect a match between observed and theoretical buoyancy gradient spectra only upon incorporating rigid-wall dissipation.\\
Fig. \ref{f1_exp}a,b present two snapshots of the observed buoyancy gradient field, $b_{\zeta}$, in steady state, with the derivative taken in the phase propagation direction of the first branch. The Fourier spectra along the depicted transects are shown in Fig. \ref{f1_exp}c,d, together with the theoretical spectra with and without rigid-wall dissipation. Fig. \ref{f1_exp}d also includes the spectrum of the numerical simulation for the large tank set-up, discussed below. \\
It is clear that for both experimental set-ups the correspondence between the observation and our 3D model is best. 
This supports our new conclusion that dissipation at the rigid walls is significant even for very small ratios of boundary layer thickness over lateral half width, $d_0/W \sim \mathcal{O}(10^{-2})$. \\

\noindent Fully 3D simulations are run for the 'large tank' set-up (see Table \ref{t1}) with the method of spectral elements, which combines the accuracy and high resolution of spectral methods with geometric flexibility of finite element methods (See \cite{Br16,Si16} for details on the numerical method). Fig. \ref{comparison_with_buoyancy} presents two snapshots of the steady state buoyancy field in a $\xi=$ constant plane (dot-dashed transect in Fig. \ref{f1_exp}d), intersecting the first wave attractor branch in the phase-propagating direction, $\zeta$. We present only $\sim 6$\% of the transversal wall-to-wall distance, to magnify the boundary layer structure near the lateral wall (here at $y=0$). For comparison, we show the theoretical buoyancy field for spectra with and without rigid-wall dissipation. The theoretical buoyancy field for the fully dissipative spectra (middle panels, max. amplitude scaled to max. amplitude of simulation) agrees with the numerical simulation remarkably well. In contrast, neglecting rigid-wall dissipation leads to a much thinner wave attractor, which might even be unstable to triadic resonance instabilities for this experiment. \\
Fig. \ref{comparison_with_buoyancy} also visualizes the complex structure of the buoyancy field in the lateral boundary layer, which is of relevance to secondary processes, such as mean flow generation. \\
Last but not least: The comparison of buoyancy gradient spectra in Fig. \ref{f1_exp}d shows that simulated and experimentally observed spectral properties agree very well, thereby confirming that wall dissipation is also important for the numerical simulation.   \\
Our results suggests that similar 2D simulations by \cite{Gr08, Sc13}, meant to replicate quasi-2D laboratory set-ups, probably miss significant dissipation at the lateral walls. We speculate that the lateral-wall dissipation shifts the on-set of triadic resonance instabilities towards stronger energy input, i.e. larger wave attractor amplitude $a$ in the experiments by \cite{Brouzet2016}. While the main conclusions by \cite{Sc13} on the on-set of triadic resonance instabilities remain intact, the forcing amplitudes for which the transition to instabilities take place might be underestimated. \\

\begin{figure}
                \centering
                \includegraphics[type=pdf,ext=.pdf,read=.pdf, width=0.55\textwidth]{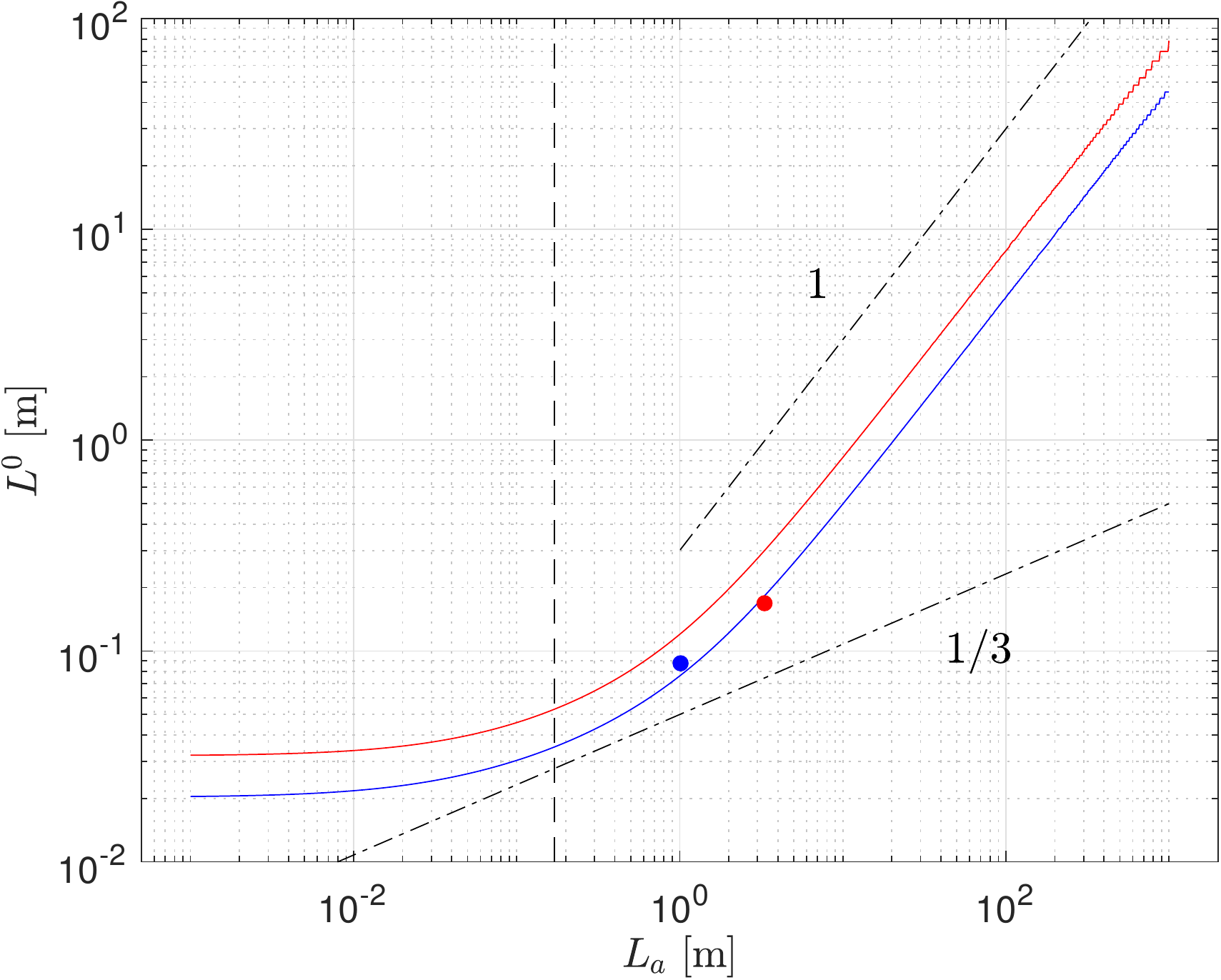}
                \caption{Theoretical wave length $L_0$, as function of orbital attractor length, $L_a$, for parameter values corresponding to 'small tank' (blue) and 'large tank' (red), with the dots showing the observed wave length at the actual orbital lengths (1 and 3 m, respectively). The vertical dashed line marked $L_a=W$; the dotted-dashed lines indicate attractor scaling for internal shear damping ($L_0^I \propto L_a^{\frac{1}{3}}$) and for lateral wall dissipation ($L_0^W \propto L_a $).  }
                \label{L0_vs_La}
\end{figure}

\noindent There is an ongoing debate on the scaling of wave attractors \citep{R01,Og05, Gr08, Ha08, Brouzet2016}. Our new analysis predicts that the scaling of wave attractors depends on the type of energy dissipation. Considering only internal shear dissipation (Eq. (\ref{Uk1}) with $P=$ constant), we get the characteristic attractor wave length $L_{0}^I=2\pi/k_{max}=2\pi \left( 3 \beta_1 \right)^{1/3} \propto \left( L_a \nu/N_0 \right)^{1/3}$, as originally found by \citet{R01} and numerically verified by \citet{Gr08}. Damping only by the lateral walls (Eq. (\ref{Uk2}) with $\beta_1=0$) results in an attractor wave length $L_0^W=2\pi \Re[\beta_2]  \propto (L_a/W)  \left(\nu/N_0 \right)^{1/2}$. Interestingly, this attractor length scale, $L_0^W$, is independent of the actual size of the 3D tank, because scaling both $L_a$ and $W$ leaves $L_0^W$ invariant. 
The dissipation at the lateral walls is negligible only if $L_0^I \gg L_0^W$, which is the case when 
$$W\gg L_a^{2/3} d_0^{1/3} \sigma_0 \frac{ [\cot \theta \ 2 (\gamma^3-1)/3]^{1/3}}{2(\gamma-1)}.$$ 
Fig. \ref{L0_vs_La} shows $L_0$ as a function of $L_a$ for the parameter values of the small and large tank set-ups by \cite{Brouzet2016}, with the dots corresponding to the observed characteristic wave lengths. The two graphs do not coincide due to slightly different parameter values, most prominently differences in angle $\alpha$ for the two set-ups. 
One can distinguish three different regimes: \\
(i) For $L_a \gg W$, lateral wall dissipation dominates, so $L_0^W \propto L_a$. \\
(ii) For $L_a \sim W$, internal shear dissipation contributes significantly, so $L_0^I \propto L_a^{\frac{1}{3}}$.\\
(iii) For very short attractors, $L_a\ll W$, dissipation at the reflecting walls dominates, so $L_0$= constant, independent of $L_a$. \\
The presented experiments fall into the transition between region (i) and (ii). This stresses the importance of previously unrecognized dissipation at rigid walls.

\section{Concluding remarks}
\label{conclusions}

From our theoretical analysis it is evident that the structure of a wave attractor in equilibrium is primarily determined by wave focusing, viscous dissipation at the rigid boundaries (mostly at the lateral walls), as well as viscous dissipation in the internal shear layers. Contrary to what was previously suggested, we show that the quasi-2D experiments by \cite{Ha08} cannot be captured by the theoretical spectrum of a 2D steady state wave attractor, which takes only internal shear dissipation into account. We close the gap between observations and theory by adding viscous dissipation at the lateral walls, which are the primary contact surfaces of the attractor and the rigid boundaries in the experiment by \cite{Ha08}. It is clear that rigid-wall dissipation also plays an important role in the experiments by \cite{Brouzet2016}. 

Contrary to previous studies, we find that the shape of the equilibrium wave attractor in the classical trapezoidal set-up is not only dependent on the properties of the stratified fluid (viscosity $\nu$, \BV frequency $N_0$), the geometry of the tank (width $W$, wave attractor length $L_a$, sloping wall angle $\alpha$) and forcing frequency $\omega_0$, but also on the nature of the energy input, which determines  the period-$1$ function $P$ in equations (\ref{Uk1}), (\ref{Uk2}) and (\ref{Uk3}). Whereas the fluid  properties and geometry determine the characteristic cross-beam wave length of the wave attractor, the nature of the energy input sets the fine structure of the equilibrium wave attractor. The role of the period-$1$ function $P$ remains vague, and more research is needed to understand the relation between a single wave number energy input and a continuous steady state wave attractor spectrum. 

In the ocean, sites where internal waves propagate parallel to a rigid vertical boundary over long distances are sparse; the channel between two coral atolls studied by \cite{Ra16} being such exceptional example. Wave beam reflection at bottom topography is much more common. To the best of our knowledge, we are the first to explicitly determine the dissipation due to such reflection. Our assumption of a stable laminar boundary layer holds in the ocean for semi-diurnal tides with amplitudes up to 32 m \citep{Bu88}. For internal tides with wave length of the order of 100 $m$ ($k_0=0.06$ rad/m), we find that the velocity amplitude decay due to non-critical reflection, $d_0 k_0 R_{\alpha}$, can amount up to $\sim 1$\%. For larger wave length, the decay is even smaller, confirming that dissipation due to laminar reflection is typically negligible in the ocean. Probably more important is the three-dimensionality of the boundary layer velocity field occurring for reflecting wave beams, which happens if incoming and outgoing beams point in different horizontal directions. It is well known that the 2D steady-state similarity linear solutions for collinear viscous wave beams by \cite{TA03} can also be valid in the nonlinear regime. This may change in the vicinity of the rigid boundary, where Reynolds stresses may become large. Consequences can be the generation of strong mean flows, such as observed in the simulations by \cite{KZS10} and by K. Raja (personal communication), or triadic resonance instability \citep{Br16a,Br17}.  Both scenarios may result in the break-down of the internal wave beam, strong energy dissipation near the reflecting boundary and potentially vertical mixing \citep{Da17}.

\section*{Acknowledgement}
We thank T. Dauxois, E. Ermanyuk, J. Frank, S. Joubaud and K. Raja for helpful discussions. INS is partially supported by the Russian Foundation for Basic Research
15-01-06363, and the Russian Science Foundation 14-50-00095. The numerical simulations were performed on the supercomputer Lomonosov of Moscow State University.

\bibliographystyle{jfm}
\bibliography{wave_attractor_damping}

\end{document}